\documentclass[10pt, english,showkeys, prl,aps,twocolumn,floatfix,superscriptaddress]{revtex4-1}

\usepackage{braket}
\usepackage{amsmath}
\usepackage{float}
\usepackage{xcolor}
\usepackage{graphicx}% Include figure files
\usepackage{dcolumn}% Align table columns on decimal point
\usepackage{bm}% bold math
\usepackage{hyperref}% add hypertext capabilities
\usepackage{microtype}
\begin{document}

\title{Finding semi-optimal measurements for entanglement detection using Autoencoder Neural Networks }

\author{Mohammad Yosefpor}
\affiliation{
 Department of Physics, Sharif University of Technology, Tehran, Iran
}
\author{Mohammad Reza Mostaan}
\affiliation{
 Department of Physics, Sharif University of Technology, Tehran, Iran
}
\author{Sadegh Raeisi}
\email{sraeisi@sharif.edu}
\affiliation{
 Department of Physics, Sharif University of Technology, Tehran, Iran
}

\begin{abstract}
Entanglement is one of the key resources of quantum information science which makes identification of entangled states essential to a wide range of quantum technologies and phenomena. 
This problem is however both computationally and experimentally challenging. 
Here we use autoencoder neural networks to find semi-optimal set of incomplete measurements that are most informative for the  detection of entangled states. We show that it is possible to find high-performance entanglement detectors with as few as three measurements. Also, with the complete information of the state, we develop a neural network that can identify all two-qubits entangled states almost perfectly.
This result paves the way for automatic development of efficient entanglement witnesses  and  entanglement detection using machine learning techniques. 
\end{abstract}


\maketitle

%%%%%%%%%%%%%%%%%%%%%%%%%%%%%%%%%%%%%%%%%%%%%%%%%%%%%%%%%%%%%%%%%%%%%%%%%%%%%%%%%

%%%%%%%%%%%%%%%%%%%%%%%%%%%%%%%%%%%%%%%%%%%%%%%%%%%%%%%%%%%%%%%%%%%%%%%%%%%%%%%%%%%%%%
Mathematical structure of quantum mechanics allows for a peculiar kind of correlation, known as 'Entanglement' that in some aspects, is more powerful than classical correlations \cite{bell1964einstein}. Entanglement is known to be one of the key resources in quantum information theory \cite{Cohen2008} that empowers many quantum technologies such as quantum metrology \cite{giovannetti2011advances}. This makes entanglement detection uniquely essential for to a variety of quantum applications \cite{guhne2009entanglement}.  
In contrast to separable states, 
Entangled states cannot be written in terms of a convex combination of the product of density matrices. This reduces the entanglement detection  to determining if a state is in the convex hull of product states \cite{horodecki1997separability}. 
This problem is also known as the 'separability problem' and  is  NP-Hard \cite{Vedral1997,gharibian_2010}.

Although exact entanglement identification for the full Hilbert space is challenging, it is possible to construct tools that can identify some but not all entangled states. 
These are known as 'entanglement witnesses'. 
Mathematically, a witness provides a sufficient but not necessary condition for entanglement. If the state satisfies this condition, it is entangled, but if it does not, the witness reveals no information with regards to the separability of the state \cite{Terhal2002, horodecki_2009}.

\begin{figure}[t]
    \centering
    \includegraphics[width=0.95\linewidth]{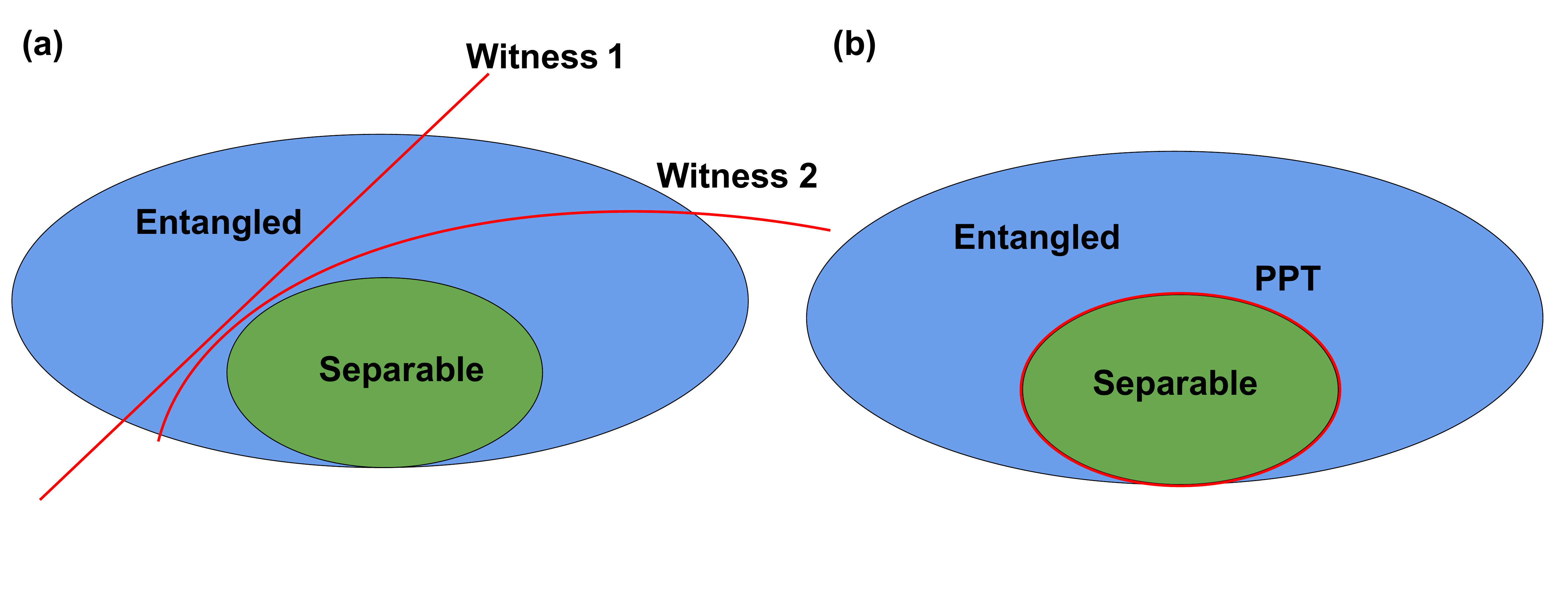}
    \caption{(a) Entanglement witnesses divide the space of states into two partitions. (b) PPT as an entanglement witness in a bi-partite system with $d_A\times d_B \leq 6$ 
(where $ d_A $ and $d_B$ are the dimensions of the Hilbert spaces of the first and second subsystems respectively)  
partitions the space into exactly the set of separable states and the set of entangled states.}
    \label{fig:witness}
\end{figure}

Figure \ref{fig:witness}(a) gives a schematic picture of how a witness identifies entangled states. It divides the Hilbert space into two partitions, one that only contains entangled states, and an other one that may include of both separable and entangled states. 

A variety of approaches have been proposed to provide some entanglement witnesses \cite{horodecki_2001,Wootters1998,Terhal2000,Rungta2001,DeVicente2007}. 
Also recently, different machine learning techniques have been developed for detection of entanglement. These include forest algorithms \cite{Wang2017}, 
neural networks \cite{Lu2018,Gray2018,Deng2018,Levine2018,Liu2018,Qiu2019},  
reinforcement  learning
and restricted Boltzmann machines\cite{Harney_2020,Gao2018,Ma2018}.
Some of these methods require the full description of the state, which in general, might require too many measurements and be infeasible experimentally. Ideally, it is favourable to  witness entanglement with as few measurements as possible. This requires finding optimal measurements that are most informative with respect to entanglement of the states. 
Note that although for the detection a general entangled states, the full density matrix (full tomography) is required, it is still possible to detect subsets of entangled states from incomplete data of a few measurements \cite{horodecki_2009}.
In this article, we propose to use autoencoder neural networks to find semi-optimal measurements for entanglement detection and construct new entanglement witnesses that can detect entangled states with incomplete data, e.g. with as few as three measurements. More specifically, our methods takes as input the constraint on the number of measurements and returns both the optimized measurements and the witness which uses the result of those optimized measurements to detect entanglement. 

Autoencoders are a type of neural network that are designed to find compressed encodings of the inputs that contain the information relevant to a specific target task. In our case, this task is the detection of entanglement and the autoencoder would find a set of few measurements that are optimized for this task. Naively, for a given number of measurements $m$, we train a neural network with a bottleneck layer of $m$ nodes and in the training process, the $m$ nodes are optimized such that the full network can, to the best of its ability, detect entangled states. This way, the autoencoder neural network finds the optimal $m$ measurements that can be used for detection of entanglement. 

Here we apply our method to the two-qubit system, but this method could be generalized for systems with higher Hilbert space dimensions. 
For the two-qubit system, 15 measurements would provide the full information of the state. With the full information, our method can identify entangled and separable states almost perfectly. As we reduce the number of measurements,  the performance of the witness decreases and it can only identify smaller portion of the entangled states. 
We also consider the situation where the state is cylindrically symmetric and find that symmetry can significantly enhance the performance of the resulting entanglement witness and even with a few measurements, it is possible to find highly accurate entanglement detectors.

%%%%%%%%%%%%%%%%%%%%%%%%%%%%%%%%%%%%%%%%%%%%%%%%%%%%%%%

The neural network witness takes states as input and indicates whether or not the state is entangled at the output. 
For the input, we need to find a representation for the state of a two-qubit system. In quantum information theory, the state of a physical system is represented by a density matrix $\rho$ which is a Hermitian, non-negative operator with trace one \cite{chuang}. We use the basis of Pauli matrices to express the density matrix and feed this representation for the input of the neural network. Mathematically that is
\begin{equation}\label{eqn:rho_2}
   \rho = \frac{1}{4}\sum_{ij} \Gamma_{ij} \sigma_i \otimes \sigma_j 
\end{equation}
where $\sigma_i$s are Pauli matrices and $\sigma_0$ denotes the identity operator $\mathcal{I}$.
$\Gamma_{ij}$s could be found as:
\begin{equation}\label{eqn:feature_cal}
    \Gamma_{ij} =  tr(\rho (\sigma_i \otimes \sigma_j))
\end{equation}

The reason we choose this basis is because $\Gamma_{ij}$ are the expectation values of $(\sigma_i \otimes \sigma_j)$ measurements on the system. So for two-qubit systems, these 15 measurements (neglecting the trivial $\Gamma_{00} = 1$ resulted from the identity operators) provide a representation of the  density matrix \cite{chuang}.

The final goal is to use the autoencoder to find fewer measurements that can be used to detect entanglement, at least for a subset of the states.

Since we use supervised learning techniques, the label of the states, i.e. whether or not they are entangled, needs to be  determined. We use the Peres-Horodecki criterion (also known as positive partial transpose or PPT criterion).
This witness states that if the partial transpose of the density matrix is negative, the state has to be entangled \cite{horodecki_2009}. 
This is because if $\rho_{AB}$ is separable then its partial transpose with respect to one party ($\rho_{AB}^{T_B}$) should be non-negative \cite{horodecki_2009}. So, if $\rho_{AB}^{T_B} < 0 $ then the state has to be entangled. 
For the special case of when the %Hilbert space 
dimensions of the subsystems satisfy  $d_A\times d_B \leq 6$ (where $ d_A $ and $d_B$ are the dimensions of the Hilbert spaces of the first and second subsystems respectively) PPT gives a necessary and sufficient condition for separability \cite{horodecki_2009}. See figure \ref{fig:witness}(b)) for a schematic picture of PPT witness. 
For two-qubit systems, Augusiak et al proposed a simplified form of the PPT criterion \cite{Augusiak2008}.
\hbadness=99999
\begin{equation}\label{eqn:augusiak}
\det(\rho_{AB}^T) \geq 0 \iff \text{separable}. 
\end{equation}
This is used to determine the labels of the density matrices.
Note that in general, PPT only gives a necessary condition for separability  which makes it challenging to extend this approach to higher dimensions. 
\begin{figure}[t!]
    \centering
    \includegraphics[width=\linewidth]{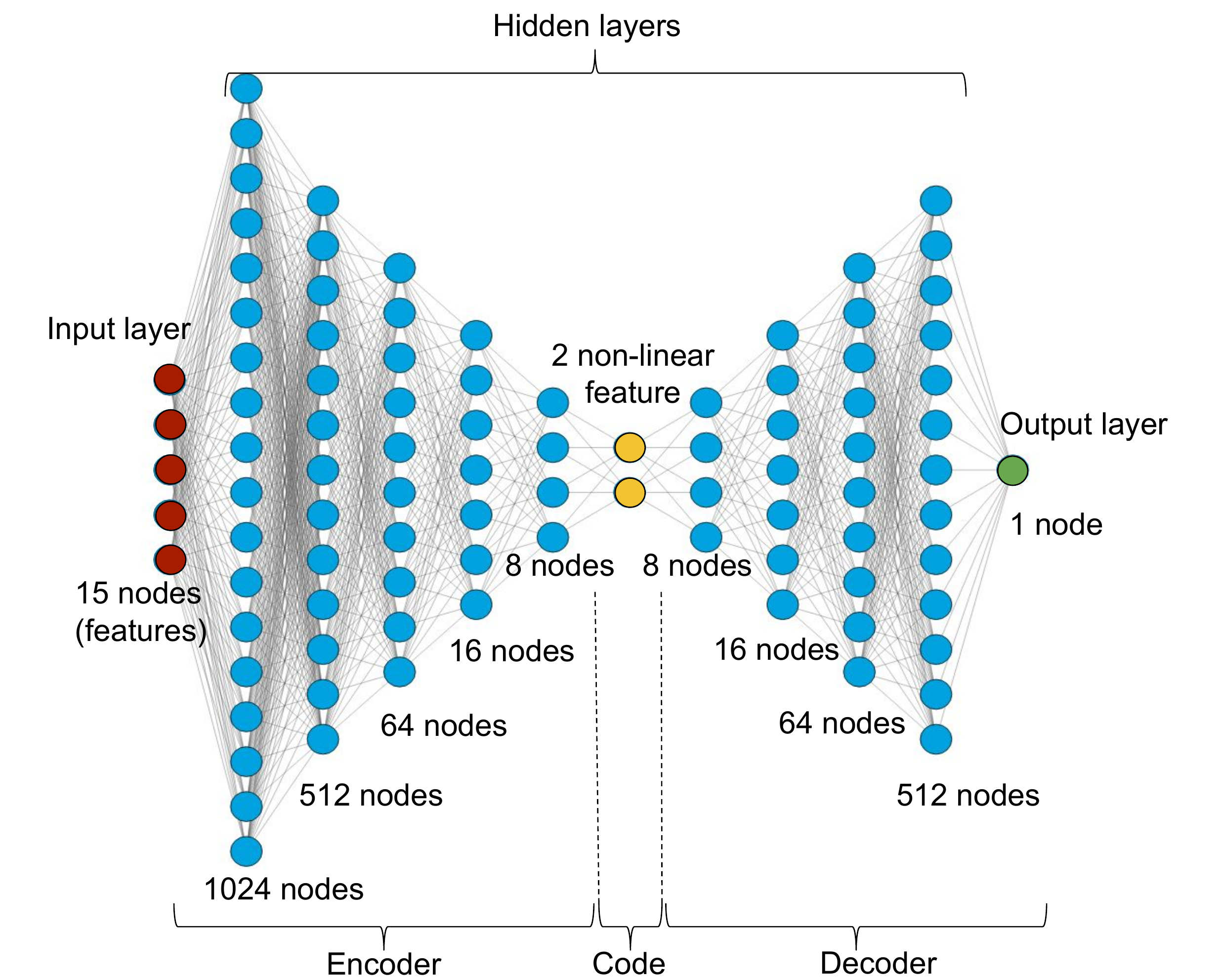}
    \caption{The architecture of the autoencoder that gives linear codes. The code layer has no activation and is a linear function of the inputs. }
    \label{fig:NN_nonlin}
\end{figure}

We use autoencoder neural network \cite{Baldi2012} to detect entangled states. The autoencoder architecture is shown in figure \ref{fig:NN_nonlin}, the number of nodes starts to decrease first and then it starts to increase before it gets to the final node. It is common to refer to the first part of the network as the 'encoder' and the last part as the 'decoder'. Also the layer with the fewest nodes is referred to as the 'code'. 
This layer, acts as a bottleneck for the flow of information. It means that if the full network can effectively detect entangled states (or a subset of them), then the information relevant to the entanglement of the states has to be coded in the code layer. 

In our case, the network is starting with the full density matrix and the encoding can be seen as finding the optimal measurement that contains the information relevant to the entanglement.

The idea is that we train this neural network and the full network gives an entanglement witness that works with complete information. But the network does more. It also finds an encoding of the state that would keep only the information relevant to the entanglement. 
In other words, we can use the encoder to find the few measurements that are most informative with regards to the entanglement and directly measure them. 

We trained a model with the architecture shown in figure \ref{fig:NN_nonlin} 
and achieved an accuracy of 98\%. The confusion matrix of the model has been demonstrated in table \ref{tab:conf_mat}.

\begin{figure}[t]
    \centering
    \includegraphics[width=\linewidth]{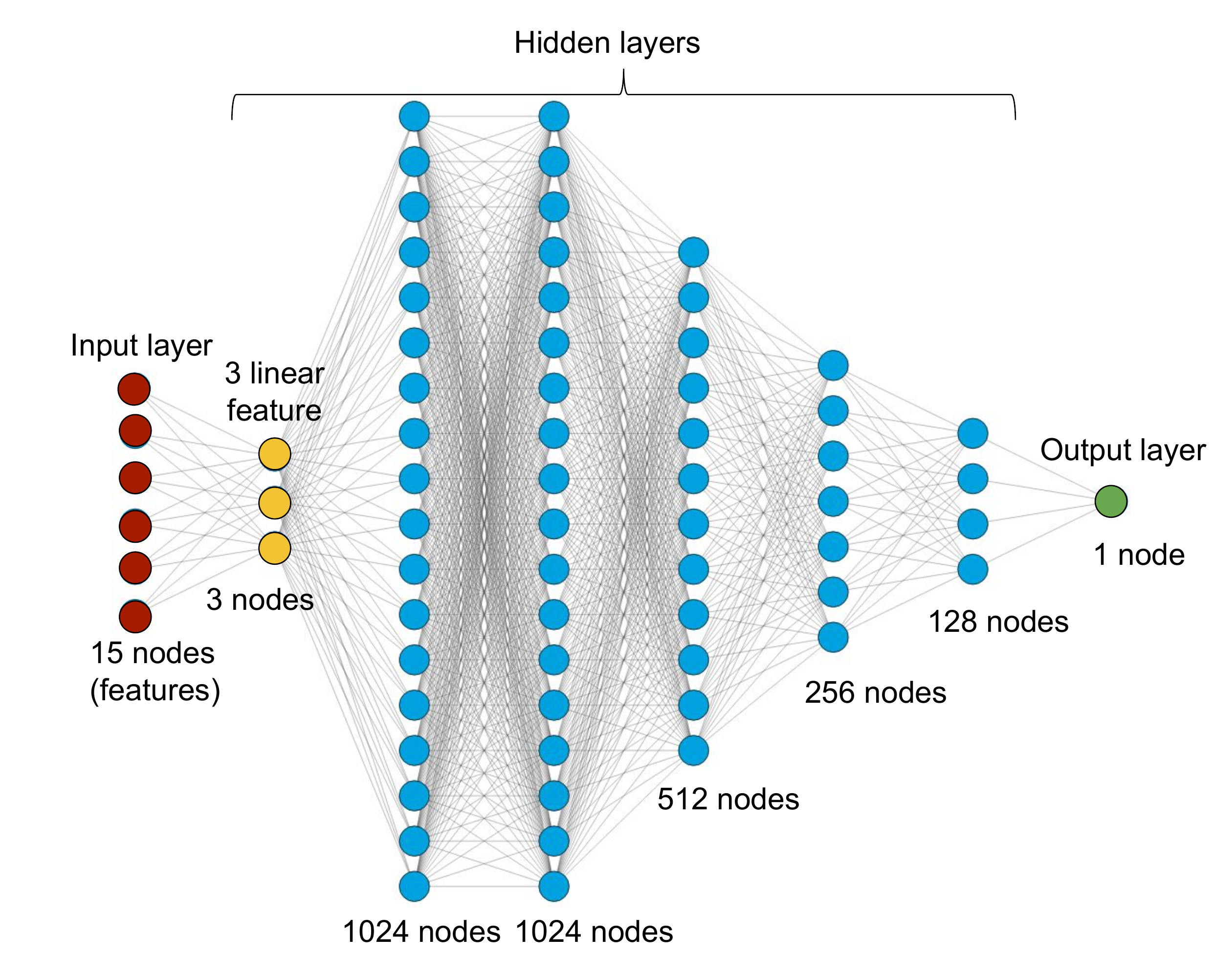}
    \caption{The architecture of the neural network with first layer consisting of only 3 nodes with no activation function which forces the network to choose 3 linear combination of the features.}
    \label{fig:NN_lin}
\end{figure}

\begin{table}[h]
    \centering
    \[
\begin{array}{c c} &

\begin{array}{c c} \text{pred. separable\quad} & \text{\quad pred. entangled} \\
\end{array} 
\\
\begin{array}{c c}
\text{true separable} \\\\\\
\text{true entangled}
\end{array} 
&
\left[
\begin{array}{c c}
\textcolor{green}{99.72\% \;(314865)}\hspace{1.2cm} & \textcolor{red}{0.28\%\;(876)}  \\\\\\
\textcolor{red}{0.16\%\;(1105)}\hspace{1.2cm} & \textcolor{green}{99.84\%\;(683154)}  
\end{array}
\right] 
\end{array}
\]
    \caption{Confusion matrix of the autoencoder neural network. The diagonal elements indicate the percentages (and the number) of separable and entangled states that are classified correctly. The top right element (in red) gives the percentage (and the number) of separable states classified as entangled and reflects the precision of the classifier. Similarly, the bottom left element gives the percentage (and the number) entangled states that were not detected by the classifier and reflects the detection power (recall) of the classifier.  }
    \label{tab:conf_mat}
\end{table}
The nodes in the code layer of this model are a non-linear function of the 15 features of the full state. This corresponds to a set of non-linear measurements on the system.
With some tuning, one may even recover the PPT criterion. But it is not really helpful because non-linear measurements are usually challenging to implement experimentally. 
So we redesign the model and look for linear measurements. That is, we want the nodes in the code layer to be linear functions of the input layer which gives the model in figure \ref{fig:NN_lin}. In this architecture, the first layer consists of only $m$ (desired number of measurements) nodes without any activation function. So the network is forced to choose $m$ linear combination of these 15 features. Each linear combination corresponds to a linear measurement.
To minimize the loss, the network changes the weights of the first layer and therefore it finds $m$ semi-optimal linear measurements which can be used to detect entanglement.

\begin{figure}[ht!]
    \centering
    \includegraphics[width=\linewidth]{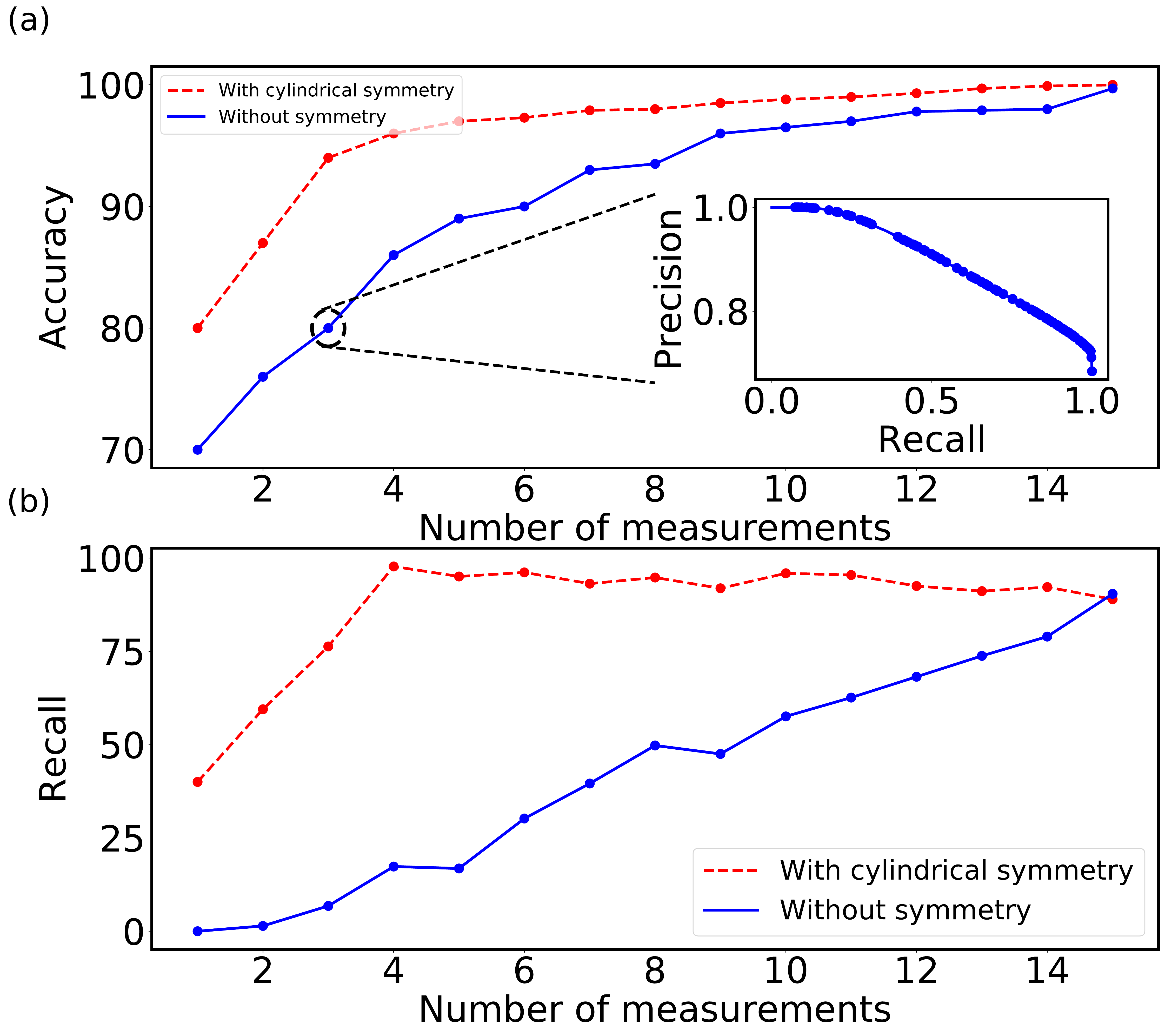}
     \caption{Accuracy and recall of prediction model for different number of measurements. As expected, the larger the $m$ the higher the accuracy and recall of the classifier. In Panel (a) the plot in blue and red are the accuracies of the models with no symmetry and cylindrical symmetry respectively. Panel (b) illustrates the recall for the models with precision one, i.e. an entanglement witness. The plot in blue and red are the recalls of the models with no symmetry and cylindrical symmetry respectively. }
    \label{fig:acc-rec-n}
\end{figure}

The model has been trained for different values of $m$ and the results are shown in figure \ref{fig:acc-rec-n}(a).
Also by modifying the threshold for the classification of the data to reach the precision of 100\%, an entanglement witness has been created, and the percentage of the entangled quantum states that these witness can detect from all of the entangled states (recall) are shown in figure \ref{fig:acc-rec-n}(b)
%[hb!]

\begin{figure}[H]
    \centering
    \includegraphics[width=\linewidth]{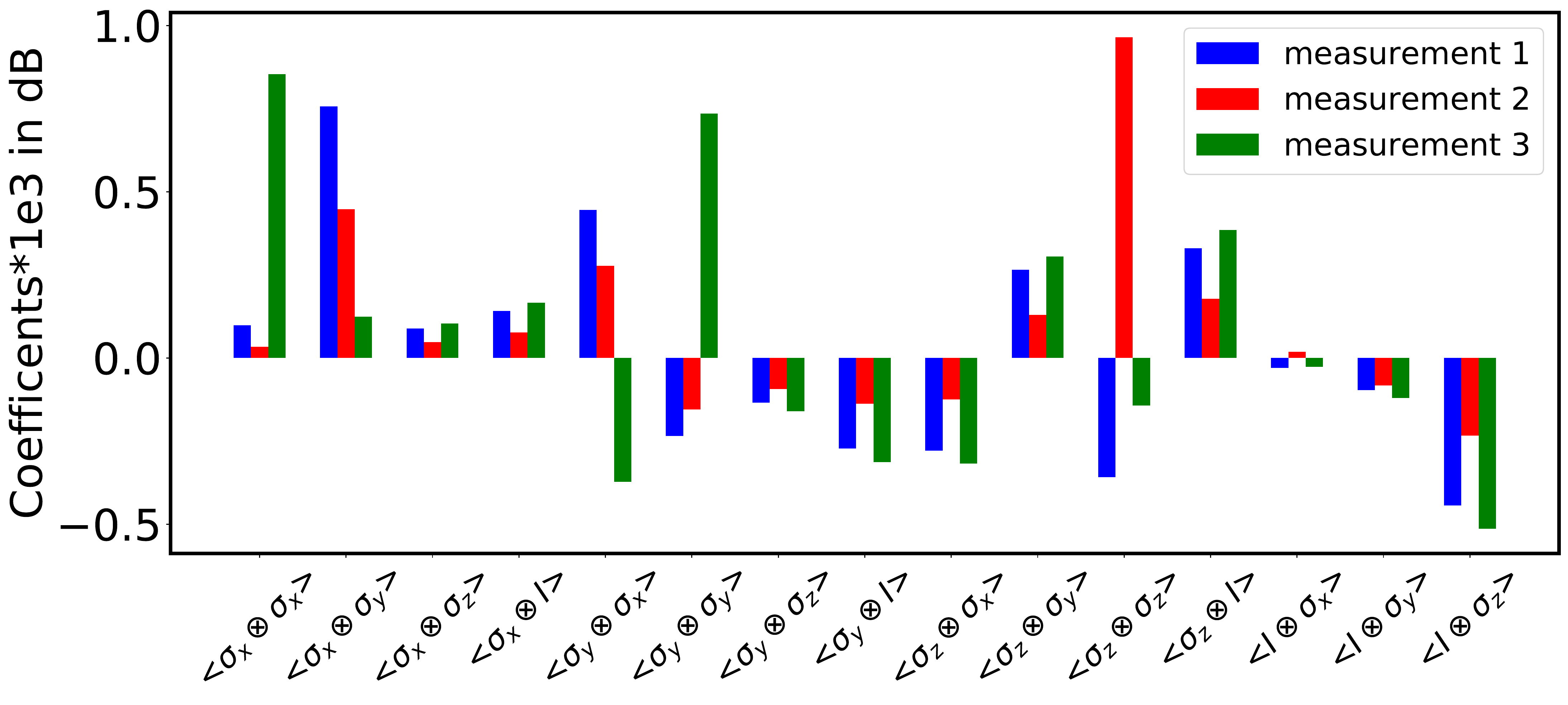}
    \caption{The coefficient of the linear code layer of Neural Network with $m=3$ nodes. These weights indicate how each of the resulting semi-optimal measurements of the autoencoder depend on the Pauli measurements.}
    \label{fig:first-layer}
\end{figure}

For example, this model achieves an accuracy of about 80\% using only three linear measurements. 
These three semi-optimal measurements can be expressed in terms of the input layer, i.e. Pauli measurements. These weights are depicted in figure \ref{fig:first-layer}. 

%%%%%%%

%%%%%%%%%%%%%%%%%%%%%%%%%%%%%%%%%%%%%%%%%%%%%%%%%%%%%%%%%%%%%%%%%%%%%%%%%%%%%%%%%
Often the quantum states prepared in the lab have some symmetries. These symmetries generally pose some constraints on the density matrix and as a result, could simplify entanglement detection. 

One of the common symmetries in state preparation is the cylindrical symmetry, i.e. 
rotations around the z-axis does not change state. 
Here, we repeat our idea with the cylindrical symmetry for the two-qubit systems. 
The results for symmetric states are shown in figure \ref{fig:acc-rec-n}. 
The model can achieve high accuracy with a few number of linear measurements in the case of cylindrical symmetry. For example, with even three measurements, it reaches 94\% of accuracy. 

In conclusion, we proposed to use autoencoder neural networks to find optimal measurements for entanglement detection and constructed entanglement detectors that, with as few as three measurements, could achieve accuracies as high as $80\%$ over all density matrices. 

On a fundamental level, our work provides a way to find a few, optimized measurements that can be most informative with respect to the entanglement detection. On a practical level, this technique can build upon existing theoretical entanglement criterion (like the PPT for two qubits) and construct entanglement witnesses that can work with significantly fewer measurements and reduce the experimental cost.

As a side result, we found an entanglement detector that with the full information of the state, can identify all entangled and separable states almost perfectly. 
For symmetric states, the performance of this technique would significantly improve. 

Our models present a proof of concept for the idea of using autoencoders for finding optimized measurement for entanglement detection. This work was limited by our computational resources and the  models can potentially be further improved by further training and by better tuning the hyper-parameters like the depth of the networks. It remains open to see how far the performance of this idea can be pushed and what its limitations are.

The idea of using autoencoders for optimization of measurements can be extended beyond two-qubit.

It would be interesting to apply our method to higher dimensional systems (beyond $2\times 2$).  
Further, this approach can be extended to other applications in quantum information \cite{zhou2020entanglement, zhou2019detecting, zhou2019decomposition}. The technical details and the codes for this work is publicly available at \cite{gitlab}.

\begin{acknowledgments}
	This work was supported by the research 
	grant system of Sharif University of Technology (G960219). %,
\end{acknowledgments}

\bibliographystyle{apsrev4-2}
\bibliography{main}

\end{document}